\documentclass{pasj00}
\draft
\begin{document}
\SetRunningHead{Kawano \& Oguri}{Time Delays for SDSS J1004+4112}
\Received{2005/10/21}
\Accepted{-/-/-}

\title{Time Delays for the Giant Quadruple Lensed Quasar SDSS
J1004+4112: Prospects for Determining \\ the Density Profile of the
Lensing Cluster}

\author{Yozo \textsc{Kawano}}
\affil{Department of Physics and Astrophysics, Nagoya University, 
Chikusa-ku, Nagoya 464-8062.}
\email{kawano@a.phys.nagoya-u.ac.jp}
\and
\author{Masamune \textsc{Oguri}}
\affil{Department of Astrophysical Sciences, Princeton University,
  Princeton, NJ 08544, USA.}
\email{oguri@astro.princeton.edu}


\maketitle

\begin{abstract}
We investigate the dependence of the time delays for the
large-separation  gravitationally lensed quasar SDSS J1004+4112  on
the inner mass profile of the lensing cluster. Adopting the mass model
whose innermost density profile is parameterized as $\rho\propto
r^{-\alpha}$, we derive a series of mass models which can fit
observational data and then compute the probability distribution
functions of time delays. We find that larger $\alpha$ has longer time
delays, longer tails at the higher end of the probability
distribution, and larger model uncertainties. The ratios of time
delays slightly depend on the slope $\alpha$. Among others,
time delays between images C and A (or B) have little dependence on
the inner slope, particularly when the time delays are short. The
dependence of time delays on $\alpha$ is well fitted by a linear form,
which reflects well-known degeneracy between the mass profile and time
delays. We perform a Monte-Carlo simulation to illustrate how well the
inner slope can be constrained from measurements of time delays. We
find that measurements of more than one time delays result in
reasonably tight constraints on the inner slope
($\sigma_{\alpha}\lesssim0.25$), while only one time delay cannot
determine the inner slope very well.  Our result indicates
that time delays indeed serve as a powerful tool to determine the mass
profile, despite the complexity of the lensing cluster.
\end{abstract}

\section{Introduction}
Recent high-resolution $N$-body simulations in the Cold Dark Matter
(CDM) universe have suggested that dark matter halos are described by
a universal mass profile (\cite{navarro96}, \yearcite{navarro97},
hereafter NFW). Higher-resolution simulations have revealed that
the innermost region may have steeper inner slopes than originally
suggested and may not be universal
(\cite{fukushige97}, \yearcite{fukushige01}, \yearcite{fukushige03};
\cite{moore99,ghigna00,jingsuto00,power03,fukushige04,hayashi04,
diemand04,tasitsiomi04,reed05}). While the accurate value of the inner 
slope and the universality remain controversial, the existence
of deep potential well at the center of dark matter halos in the CDM
model appears quite robust. Here the important fact is that the inner
profile of dark halos is extremely sensitive to nature of dark matter:
For instance, if one introduces self-interacting cross sections of
dark matter particles, the central density profile becomes much
shallower than expected in the standard CDM universe \citep{spergel00}.
Therefore it serves as a powerful test of collisionless CDM
paradigm. Such observational studies to constrain parameters of 
the density profile have been conducted. For instance, assuming the
hydrostatic equilibrium the mass profile of a dark matter halo can be
evaluated from the gas density and the temperature profiles of
intra-cluster medium (e.g., \cite{sato00}); weak lensing
reconstruction can reproduce the mass profile of a cluster from
ellipticities of galaxies behind the cluster
\citep{kaisq93,schneider05}. In particular, rotation
curve observations of dark-matter-dominated dwarf and low surface
brightness disk galaxies have indicated that they favor mass profiles
with a flat density core, being inconsistent with the NFW profile
\citep{salubu00,deblok02,swaters03,gentile04}. However, this apparent
discrepancy might be because of bias in gas rotation speed
\citep{hayashi04} or the destruction of central cusps by the formation
of a primordial bar \citep{weinberg02}. Thus, investigations
of the halo density profiles in different systems has been one of
the most essential issues.

Among others, one of the most promising methods to test the NFW
profile would be strong lensing by clusters of galaxies. Gravitational
lensing is unique in the sense that it allows one to detect the mass
distribution {\it directly}
(\cite{refsdal64}; \cite{bourassa73}, \yearcite{bourassa75},
\yearcite{bourassa76}; \cite{wambsganss94,keetonks97,mao98,chiba02,
kawano04,kocda04}). In particular, strong lensing by clusters
is a indispensable tool to probe the innermost region of dark matter
halos, since most baryons in clusters remain hot and diffuse and
therefore the density profile of clusters can be well approximated
by that of dark halos seen in $N$-body simulations. 
For instance, giant arcs of background galaxies with different redshifts
(Einstein radii) can strongly constrain the mass profiles of clusters
\citep{broadhurst05}. This is because the mass inside the Einstein
ring with radius $r_{\rm E}$ at which arcs build up is proportional
to $r_{\rm E}^2$ and thus multiple arcs with different redshifts
determine the masses within the different radii, equivalently the mass
slope. Multiply 
imaged quasars due to clusters of galaxies offer another direct probe of
the mass profiles of lensing clusters. Indeed, they have several
advantages over giant arcs. First, for lensed quasar systems it is easier
to correct selection bias of lensing clusters (i.e., difference between
lensed and unlensed cluster populations; see e.g., \cite{hennawi05})
because of well-known source population of quasars. Second, we can
measure time delays for lensed quasars, which serves as additional
strong constraints on the mass distribution. 

Recently, the first example of such quasar-cluster lens system, SDSS
J1004+4112, has been discovered \citep{inada03,oguri04} in the Sloan
Digital Sky Survey (SDSS; \cite{york00}). It has an unusual separation
of $\sim 15$ arcsec, which is more than twice larger than the second
largest lens Q 0957+561 \citep{walsh79}. The quasar and lensing cluster
have redshifts of $1.74$ and $0.68$, respectively. The central region
of the lens cluster is dominated by the brightest cluster galaxy. The
fifth image of the lensed quasar, which constrains inner mass
distribution of the the brightest cluster galaxy strongly, was
discovered by \citet{inada05}.

How did the lens system SDSS J1004+4112 constrain the inner mass
distribution of the dark matter halo? \citet{oguri04} modeled the lens
with singular isothermal ellipsoid (SIED) plus NFW, and found that
the image configurations and fluxes are reproduced well. More
generally, \citet{willisaha04} adopted a free-form mass reconstruction  
technique \citep{sahawi97} and pointed out that the quadruple images
alone do not give useful constraints on the inner slope of the dark
matter halo. Therefore, we need additional information: One such
information comes from arcs of background galaxies which have already
observed by \citet{sharon05}. Arcs at several radii (redshifts) can
constrain the inner slope since we can estimate the mass inside a arc
from the the arc position and shape. Another information, which is
unique for lensed quasar systems, is time delays between each images. 
It has been shown that the Hubble constant and the slope of the
density profile are both sensitive to time delays 
\citep{wambsganss94,oguri02,wucknitz02,kochanek02,ogukawa03,kochanek05},
therefore by assuming the value of the global Hubble constant,
which is now determined with better than 10\% accuracy
\citep{freedman01,spergel03}, 
we will be able to obtain useful information on the mass profile.
The image configuration of the system is very similar to that of PG
1115+080 \citep{weymann80}, however the separations are scaled by
a factor of $\sim 8$ and hence the time delays are longer by a factor
of $\sim 8^2$, which means that the shortest and longest time delays
would be $\sim 10$ days and $\sim 1000$ days, respectively.
However, the complexity of the cluster mass distribution makes the
possible range of time delays quite large \citep{oguri04,willisaha04}.
Therefore it is not obvious whether time delays can really give
meaningful constraints on the inner profile of the dark matter halo.
In this paper, we investigate in detail the capability of measuring
time delays and constraining the inner mass slope of the cluster.

This paper is organized as follows. In section \ref{sec:model} we
describe our model to predict time delays for SDSS J1004+4112. Section
\ref{sec:predtij} gives the predictions of time delays and the
probability distributions of them. Then, a likelihood analysis is
performed in section \ref{sec:const} to see how time delays constrain
the radial slope of the cluster mass. We finally discuss the results
and give conclusions in section \ref{sec:discon}. 

\section{Lens Mass Modeling\label{sec:model}}
A source at ${\bf y}$ and an $i$-th image at ${\bf x}_i$ are related
through the lens equation
\begin{equation}
 {\bf y} = {\bf x}_i - {\bf \nabla} \psi ({\bf x}_i),
\end{equation}
where $\psi ({\bf x}_i)$ is the projected lens potential
\citep{schneider92}. The vectors are defined on the sky (lens and
source planes). The lens potential can be expressed as
\begin{equation}
\psi({\bf x})=\frac{1}{\pi} \int d^2 x' \kappa ({\bf x}')
\ln |{\bf x}-{\bf x}'|,
\end{equation}
where $\kappa ({\bf x}_i)$ is the dimensionless surface mass density,
so called convergence. More specifically, the convergence is
proportional to the surface mass density $\Sigma ({\bf x})$ as
\begin{equation}
\kappa ({\bf x})=\frac{\Sigma ({\bf x})}{\Sigma_{\rm cr}},
\end{equation}
where $\Sigma_{\rm cr}$ is the critical surface mass density defined
by
\begin{equation}
\Sigma_{\rm cr}\equiv\frac{c^2}{4\pi G}
\frac{D_{\rm s}}{D_{\rm d}D_{\rm ds}},
\end{equation}
with $D_{\rm d}$, $D_{\rm s}$ and $D_{\rm ds}$ being the angular
diameter distances to the lens, to the source, and from the lens to
the source, respectively. The gravitational lenses magnify the flux of
$i$-th image by the factor of $\mu ({\bf x}_i)$:
\begin{equation}
 \mu ({\bf x}_i)=\left| {\rm det}\left(
\frac{\partial {\bf y}}{\partial {\bf x}}
\right)_{{\bf x}={\bf x}_i}\right|^{-1}.
\end{equation}
The relative time delay between $i$-th and $j$-th images is then
calculated from 
\begin{equation}
\Delta t_{ij}=\frac{1+z_{\rm d}}{c}\frac{D_{\rm d}D_{\rm s}}{D_{\rm ds}}
\left[ \frac{1}{2} \Delta {\bf x}_i^2-\frac{1}{2} \Delta {\bf x}_j^2
-\psi({\bf x}_i)+\psi({\bf x}_j)\right],
\label{eq:tij}
\end{equation}
where $\Delta {\bf x}_i={\bf x}_i-{\bf y}$. Provided that one knows
redshifts of the source and the lens, position and flux of the source,  
and mass distribution (potential) of the lens as well as the
cosmological parameters, one can calculate observable quantities such as
the number of images, image positions, fluxes, and time
delays. Turning the problem around, by measuring the redshifts, image
positions and fluxes, and relative time delays, one can constrain the
mass distribution of the lens. 

In this paper, we consider a specific quadruple lens system, SDSS
J1004+4112, the first quasar multiply lensed by a central part of a
massive cluster. There has been a variety of studies to investigate the
structures of the lensing cluster. \citet{oguri04} firstly performed  
an enormous amount of modeling of the system with realistic
two-component models. Provided that the galaxy has a
SIED and the cluster has an elliptical version
of NFW profile with an external shear, respectively, they concluded
that: 1) there is a offset between the centers of the brightest galaxy
and the cluster; 2) a wide range of the models can reproduce the
position angle of the galaxy well; 3) the elongation of modeled
cluster are also similar to the observed distribution of cluster
galaxies; 4) they found a large tidal shear ($\sim 0.2$) which
suggests significant 
substructure in the cluster; 5) there is an enormous uncertainty in
the predicted time delays between the quasar images, which indicates
that measuring the delays would greatly improve constraints on the
models; 6) measuring the time delay between image A and B would
determine the temporal ordering such as C-B-A-D and give an expected
value of the delay between image C and D; 7) the scale length of the
dark matter of the cluster would be $r_{\rm s}\ge30\arcsec$ from
predicted relation between $r_{\rm s}$ and the lensing strength (see
below for the definition of $r_{\rm s}$).
On the other hand, \citet{willisaha04} presented free-form
reconstructions of the lens with constraints of the image positions
and physical conditions. The modeling gave some important results: 1)
the projected cluster mass profile is consistent with being
$r^{-0.3...-0.5}$, which can be fitted with either the NFW or a flat
core model; 2) the residual mass maps created by subtracting the
circularly averaged surface mass density shows the significant
substructures; 3) D-A-B-C time ordering results in shift $\simeq
3\arcsec$ of the lens center and then tends to allow a lot of models
with spurious extra images --- the time ordering seems very unlikely;
4) a measurement of time delay between A and B would serve as a test
of the shallow mass profile. \citet{inada05} discovered the fifth
central image and tested the ability of the image to constrain the
mass profile of the bright cluster galaxy. They assumed a power-law
density profile $\rho(r)\propto r^{-\gamma}$ which potential
was set to have an elliptical symmetry for the galaxy and concluded
that the central bright cluster galaxy cannot have steeper mass
profile than isothermal ($\gamma \le 2$) to reproduce the flux.
\citet{sharon05} discovered multiply imaged arcs of three galaxies
at high redshifts. Their different redshifts correspond to different
radii of Einstein rings because of different critical surface mass
densities. Since multiply imaged arcs constrain the mass inside the
arcs, they obtained the average projected surface mass slope
consistent with NFW ($\sim -0.5$) from their simple (preliminary)
one-component modeling of the dark matter halo. We note that the
lensing probability can constrain the mass profile as well
\citep{oguri04,oguri04b}, but the constraint degenerates strongly with 
cosmological parameters.
 
Following \citet{oguri04}, in this paper we consider the two-component
model that consists of the brightest cluster galaxy and dark
matter components. However, an important extension of the model in
this paper is that we allow various inner slopes of dark matter
components to study their effects on time delays. More specifically, we
model the brightest cluster galaxy, which \citet{oguri04} called G1,
as an SIED
\begin{equation}
\kappa(x,y) = \frac{b}{2 \xi},
\end{equation}
where $\xi=\sqrt{x^2+y^2/q^2}$, $b$ is a characteristic deflection
angle which is related to the velocity dispersion $\sigma_v$ of the
galaxy by 
\begin{equation}
b=4\pi \frac{D_{\rm ds}}{D_{\rm s}} \left(\frac{\sigma_v}{c}\right)^2,
\label{eq:bv}
\end{equation}
and $q$ is an axis ratio of the projected mass distribution.
The separation of the lensed images is so large that the small
deviation from the isothermal profile have little effect on
predictions of the time delays. The cluster of galaxies is modeled as
a generalized NFW profile (GNFW)
\begin{equation}
\rho = \frac{\rho_{\rm s}}{(\sqrt{\xi^2+z^2}/r_{\rm s})^{\alpha}
(1+\sqrt{\xi^2+z^2}/r_{\rm s})^{3-\alpha}},
\label{eq:gnfw}
\end{equation}
where $r_{\rm s}$ is a scale length, $\rho_{\rm s}$ is a
characteristic density (computed from the scale length $r_{\rm s}$ and
the virial mass $M_{\rm vir}$), and $\alpha$ is an inner density slope 
($\rho \propto r^{-\alpha}$). The surface mass density of this GNFW
profile is obtained by integrating equation (\ref{eq:gnfw}) over $z$. 
As done by \citet{oguri04}, an external shear is added in order to
approximately include the effects of possible complex structure in the
outer region of the cluster.  

\begin{figure}
  \begin{center}
    \FigureFile(120mm,80mm){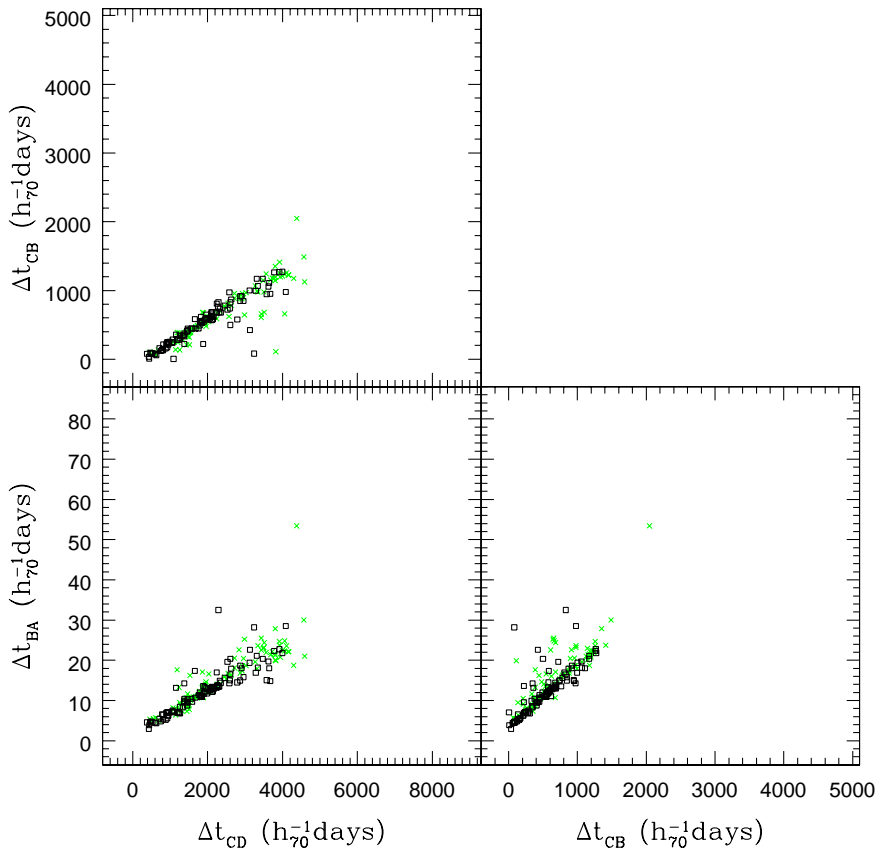}
  \end{center}
  \caption{Predictions of the three time delays for
    models with different scale length. We fix the inner density slope
    to $\alpha=1$. Green crosses show predictions for $r_{\rm
    s}=40\arcsec$ and black squares denote those for $r_{\rm
    s}=60\arcsec$.} 
\label{tij_rs40rs60al1.00}
\end{figure}

\begin{table}
\begin{center}
\caption{Constraints on Mass Models}\label{tab:pofl}
    \begin{tabular}{lcccc}
\hline \hline
     Object & $x$ (arcsec)\footnotemark[$*$] & $y$ (arcsec)
\footnotemark[$*$] &
   Flux (arbitrary)\footnotemark[$\dagger$] & P.A. (deg)
\footnotemark[$\ddagger$] \\
\hline
     A & $0.000\pm0.001$ & $0.000\pm0.001$ & $1.000\pm0.200$ & ...  \\
     B & $-1.317\pm0.002$ & $3.532\pm0.002$ & $0.732\pm0.146$ & ...  \\
     C & $11.039\pm0.002$ & $-4.492\pm0.002$ & $0.346\pm0.069$ & ...  \\
     D & $8.399\pm0.004$ & $9.707\pm0.004$ & $0.207\pm0.041$ & ...  \\
     G1 & $7.114\pm0.030$ & $4.409\pm0.030$ & ... & $-19.9\pm20.0$  \\
\hline
    \multicolumn{4}{@{}l@{}}{\hbox to 0pt{\parbox{180mm}{\footnotesize
      \footnotemark[$*$] The positive directions of $x$ and $y$ are
     defined by West and North, respectively. 
      \par\noindent
      \footnotemark[$\dagger$] Errors are broadened to 20\% to account
     for possible systematic effects. 
      \par\noindent
      \footnotemark[$\ddagger$] Degree measured east of north.
      }\hss}}
    \end{tabular}
\end{center}
\end{table}
In order to exclude unphysical situations, we assume that $b$ is
smaller than $2.25$ arcsec because any galaxies should have the
velocity dispersion smaller than $400 {\rm km s}^{-1}$. The inner mass
slope $\alpha$ is restricted to $0.5$, $0.75$, $1.0$, $1.25$, $1.5$
mainly because of the computational limitation. We note that
NFW originally proposed $\alpha=1$ universal model and recent N-body
simulations have suggested that $\alpha$ is approximately between 1 and
1.5. We fix $r_{\rm s}=40\arcsec$, because \citet{oguri04} concluded
that $r_{\rm s} \geq 30\arcsec$, which is much larger than the values
of the distances between the lens center and the images. In an inner
region ($r\ll r_{\rm s}$), profiles with different scale lengths 
are not much different from each other while varying $\alpha$ has much
larger effects. This is clearly demonstrated in Figure
\ref{tij_rs40rs60al1.00},
where we show the predicted time delays of $r_{\rm s}=40\arcsec$,
$60\arcsec$ (each point in the Figure corresponds to one of
statistically acceptable models; see below for more details). This
figure implies that the dependence on the scale length 
is not important in the case of SDSS J1004+4112, as long as the scale
length is reasonably large, $r_{\rm s}\geq30\arcsec$. 
Therefore, the number of the model parameters for each $\alpha$ is
$15$: the position, mass, ellipticity, and position angle of
the central galaxy G1; the position, mass, ellipticity, and position
angle of the cluster (GNFW); the amplitude and position angle of the
external shear; the position and flux of the source.

For observed positions and fluxes of the quasar and the central
galaxy, we use the data of \citet{inada05}, which is summarized in
Table \ref{tab:pofl}. For the purpose of investigating relative time
delays between the large-separation images, information of the 
fifth image is not included in our calculation, because its position
and flux are supposed to be governed mainly by the detailed 
mass distribution of the bright cluster galaxy G1 rather than that of
the dark matter halo. Actually, we assume the galaxy to have an
isothermal mass profile and then our profile has passed the test of
\citet{inada05}. In addition to image positions and fluxes of the
the quasar and the galaxy, the data of a position angle of the galaxy
($\theta_g\sim -19\arcdeg.9 \pm 20\arcdeg.0$) is used to constrain our
model since the projected mass and light are generally aligned to each
other \citep{keeton98}. As a result, there are $15$ constraints on mass
models.

Since the numbers of the model parameters and the observables are
equal to each other, it is expected that a wide range of parameter
space can fit the data (e.g., \cite{keetonw03,oguri04}). Following their work,
we derive a series of acceptable models as follows. First, we start
from a random starting point in the parameter space.
We put the parameters initially on the range $0\arcsec<b<2\arcsec.25$,
$6\arcsec.964 \le x({\rm G1})\le 7\arcsec.264$, $4\arcsec.259\le
y({\rm G1})\le 4\arcsec.559$, $0\le e_{\rm g}\le 0.9$, $-59\arcdeg.9\le
\theta_{\rm g}\le 20\arcdeg.1$, $0<\kappa_{\rm s}=\rho_{\rm s} r_s/
\Sigma_{\rm cr}\le 1.5$, $0\le e_{\rm c}\le 0.9$, $-90\arcdeg.0\le
\theta_{\rm c}\le 90\arcdeg.0$, $0\le \gamma \le 0.8$, $-90\arcdeg.0\le
\theta_{\gamma}\le 90\arcdeg.0$, where
$e_{\rm g}$ is the ellipticity of the galaxy, $e_{\rm c}$ and
$\theta_{\rm c}$ are the ellipticity and its position angle of the
cluster, and $\gamma$ and $\theta_{\gamma}$ are the amplitude and its
position angle of the external shear, respectively. We then perform a
$\chi^2$ minimization and find a local minimum in the $\chi^2$
surface. We adopt {\it lensmodel} package \citep{keeton01} to solve
the lens equation and to perform $\chi^2$ minimization.
If the $\chi^2$ of the minimum is $<11.8$, we regard it as an
acceptable model, and pick it up\footnote{The value of 11.8 represents
the $3 \sigma$ limit in a projected two-dimentional parameter space,
when $\chi^2_{\rm min}=0$. We note that the value was also adopted in
\cite{oguri04}.}. Any model with $b > 2\arcsec.25$ or 
unusually large ellipticities $e=1-q>0.9$ is excluded so as not to 
include unphysical models in our analysis. Because of the result 3) of
\citet{willisaha04} described above, we allow only the models that
predict C-B-A-D time ordering. By repeating this process from random
numerous starting points, we obtain a group of acceptable models. Most
($>90\%$) of local minima in the parameter space are excluded because
they are not physical or give bad fits statistically. For each
$\alpha$, we determine 100 acceptable models accordingly.

Throughout the paper we adopt a flat lambda-dominated universe with
$(\Omega_0, \Omega_{\Lambda}) = (0.3, 0.7)$, where $\Omega_0$ is the
density parameter  of matter and  $\Omega_{\Lambda}$ is the
dimensionless cosmological constant. However, our predicted time
delays can be converted to those in any other cosmologies via equation
(\ref{eq:tij}). The Hubble constant in units of $70$ km ${\rm s}^{-1}$
${\rm Mpc}^{-1}$ is denoted by $h_{70}$. 

\begin{figure}
  \begin{center}
    \FigureFile(80mm,80mm){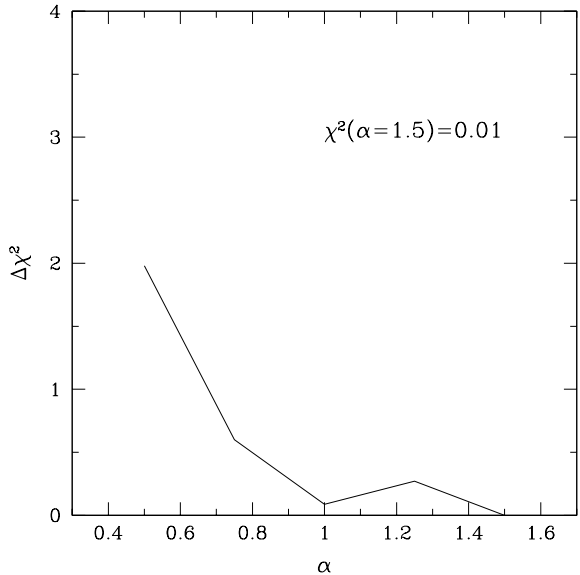}
  \end{center}
  \caption{The $\Delta \chi^2$ distribution as a function of the inner
    slope $\alpha$. The minimum $\chi^2$ was achieved at $\alpha=1.5$,
    and the value was $\chi^2=0.01$.}
\label{deltachisqalpha_rs40}
\end{figure}
\section{Predictions of Time Delays\label{sec:predtij}}
Using the method described in the previous section, we fit the
gravitational lens system SDSS J1004+4112 with varying inner mass
slopes, $\alpha=0.5$, $0.75$, $1.0$, $1.25$, $1.5$.
First, we regard $\alpha$ as a parameter and compute
$\Delta\chi^2\equiv \chi^2-\chi^2_{\rm min}$. For each $\alpha$, the
other model parameters are optimized. The result is shown in Figure 
\ref{deltachisqalpha_rs40}.
The minimum $\chi^2$ is very small because the degree of freedom
(hereafter DOF) is 0 and therefore the mass model that reproduces the
observables perfectly can exist, as is often the case with the
analysis of strong lens systems (e.g., \cite{keetonw03,pindor05}). 
This confirms earlier claims that only image positions and fluxes
cannot constrain the inner mass slope $\alpha$ very well.

\begin{figure}
  \begin{center}
    \FigureFile(100mm,80mm){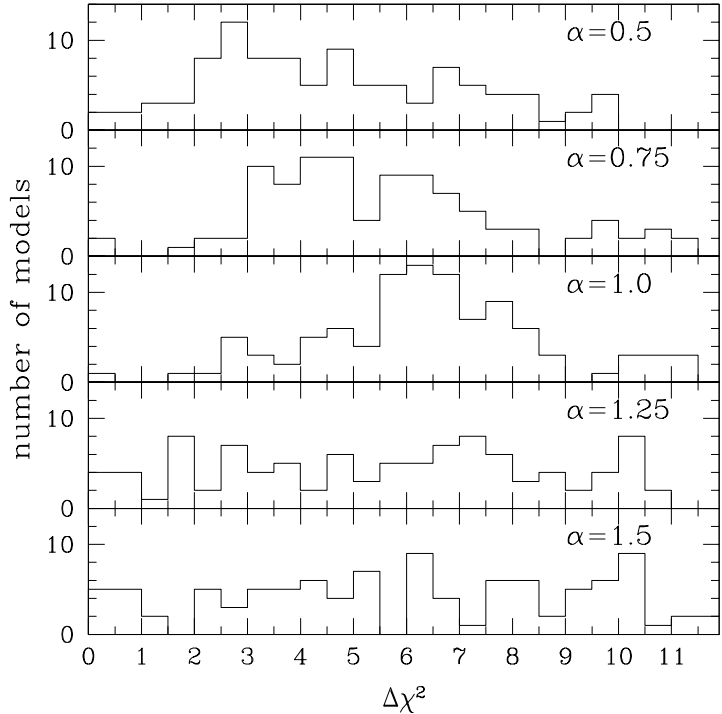}
  \end{center}
  \caption{The $\Delta \chi^2$ distributions of our 100 acceptable
  models for each $\alpha$. }
\label{deltachisqdpdalpha_rs40}
\end{figure}

\begin{figure}
  \begin{center}
    \FigureFile(120mm,80mm){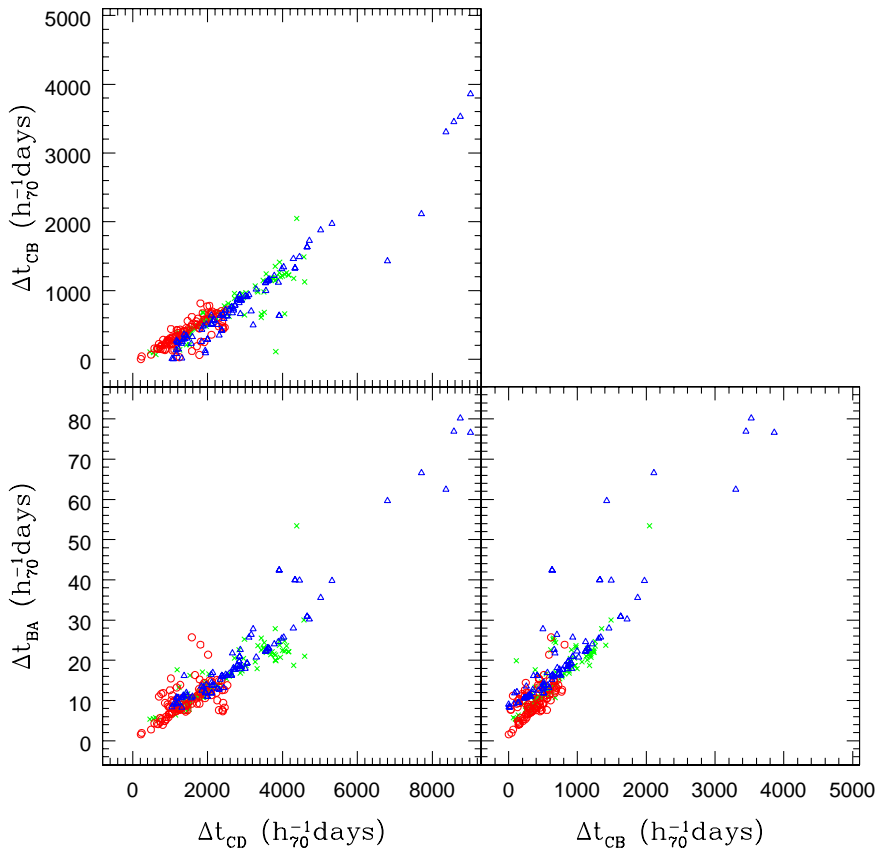}
  \end{center}
  \caption{Predictions of the three time delays for different inner
  slopes: $\alpha=0.5$ (red circles), $1.0$ (green crosses), $1.5$ (blue
  triangles). We show time delay predictions for a group of acceptable
  models: Each point indicates predicted time delays of one of the
  models.}  
\label{tij_rs40al}
\end{figure}

Figure \ref{deltachisqdpdalpha_rs40} shows the $\Delta \chi^2$ of the
models for each $\alpha$. Their distribution have no distinct feature
and therefore local minima of $\chi^2$ are expected to exist in
a broad range of the parameter space. Thus the predicted time delays
should also have a broad range of values. In what follows we consider
only local minima and neglect the distributions of $\chi^2$ around
local minima, because the differences of time delays between
different local minima are typically much larger than the
uncertainties of time delays around local minima (see also
\cite{keetonw03}). The time delays that local minima predict are shown
in Figure \ref{tij_rs40al} for $\alpha = 0.5$, $1.0$, $1.5$. As shown
in \authorcite{oguri04} (\yearcite{oguri04}, for $\alpha=1.0$), all the
two of the time delays, for instance the time delay between A and B,
and C and D, are approximately proportional to each other. We find
that this is also the case for $\alpha\neq 1$.  However, the
dispersion is so large --- one short time delay does not always give
the other two short time delays. The distribution in longer time
delays is sparse relative to that in shorter ones. As
increasing $\alpha$, the distribution shifts to longer time delay
space and the dispersion becomes larger. There is different trends in
the short time delay edges in the time delay space for different
values of $\alpha$. Thus, the ratios between time delays are expected
to vary by the inner mass slope $\alpha$. 

The tendency of increasing time delays with increasing inner slope
$\alpha$ may be explained in terms of the famous radial mass index
versus the Hubble constant degeneracy
\citep{wambsganss94,oguri02,wucknitz02,kochanek02,ogukawa03,kochanek05}.
\citet{wucknitz02} showed that modeling a lens galaxy with observed
time delay as having convergence $\kappa \propto r^{\beta-2}$ results
in the degeneracy $h_{70} \propto (2-\beta)$. More properly, larger
$\beta$ leads to smaller $h_{70} \Delta t_{ij}$ when all the images
exist at similar radii. The inner mass distribution of the GNFW
profile is $r^{-\alpha}$, hence its projected mass profile is
approximately $\propto r^{-\alpha+1}$ in the case of the GNFW
profile. Thus, larger $\alpha$ leads to longer $\Delta t_{ij}$
for the fixed Hubble constant. We confirm that this tendency, at least
qualitatively, exists even in SDSS J1004+4112 for which the existence
of the central galaxy G1 and substructures in clusters complicates the
total mass distribution. 

\begin{figure}
  \begin{center}
    \FigureFile(120mm,80mm){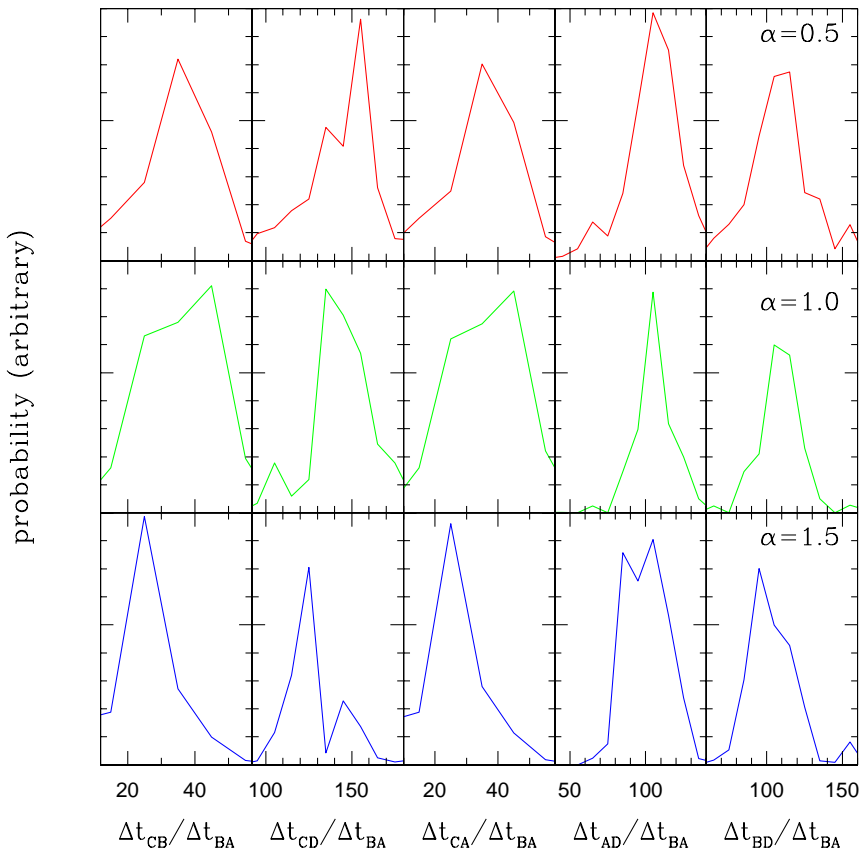}
  \end{center}
  \caption{Probability distribution functions of time delay ratios
for inner slopes $\alpha=0.5$ (red lines), $1.0$ (green lines), $1.5$
(blue lines).  Their normalizations are arbitrary.}
\label{ratiotpdf_rs40}
\end{figure}

\begin{figure}
  \begin{center}
    \FigureFile(65mm,65mm){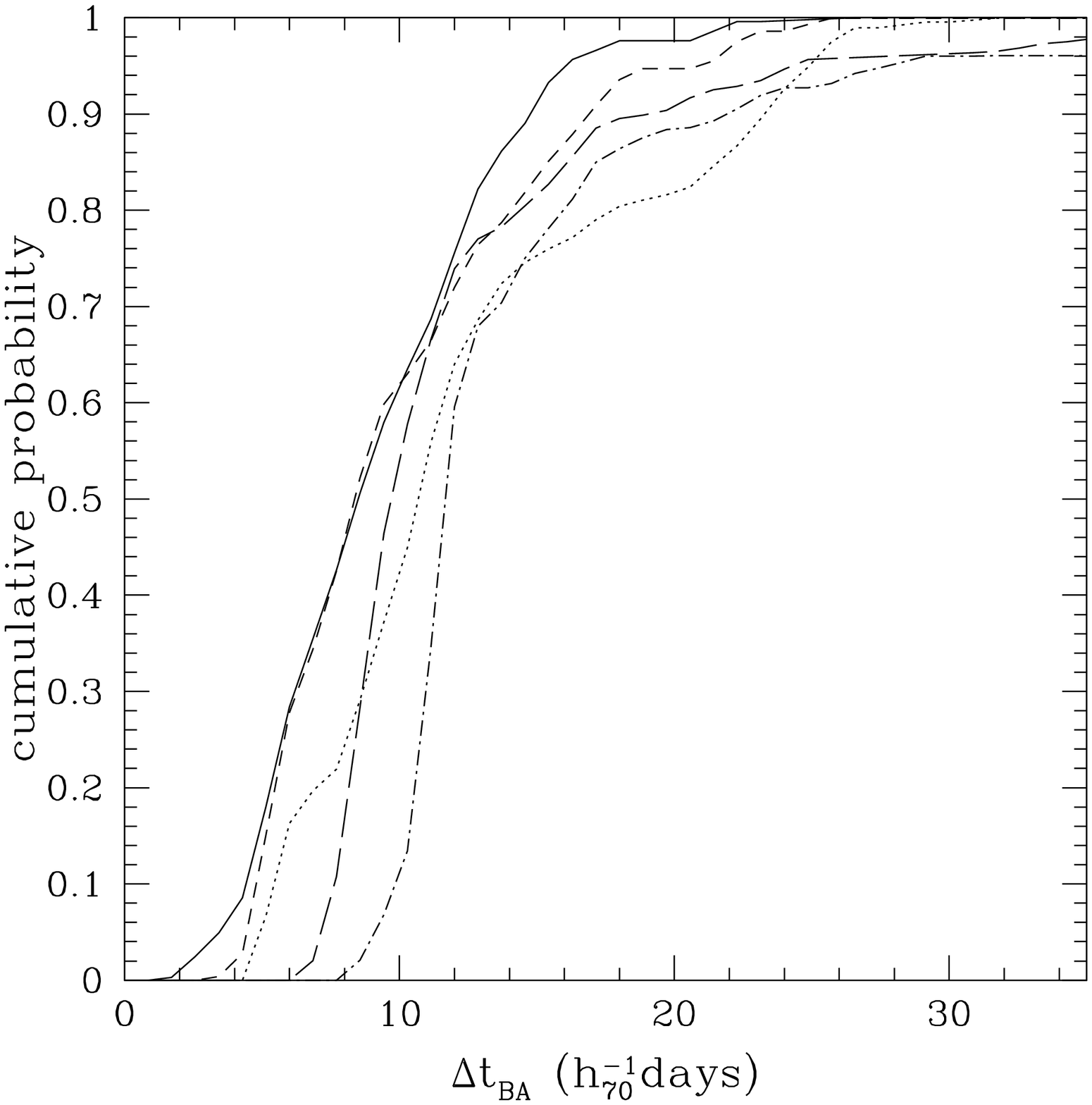}
    \FigureFile(65mm,65mm){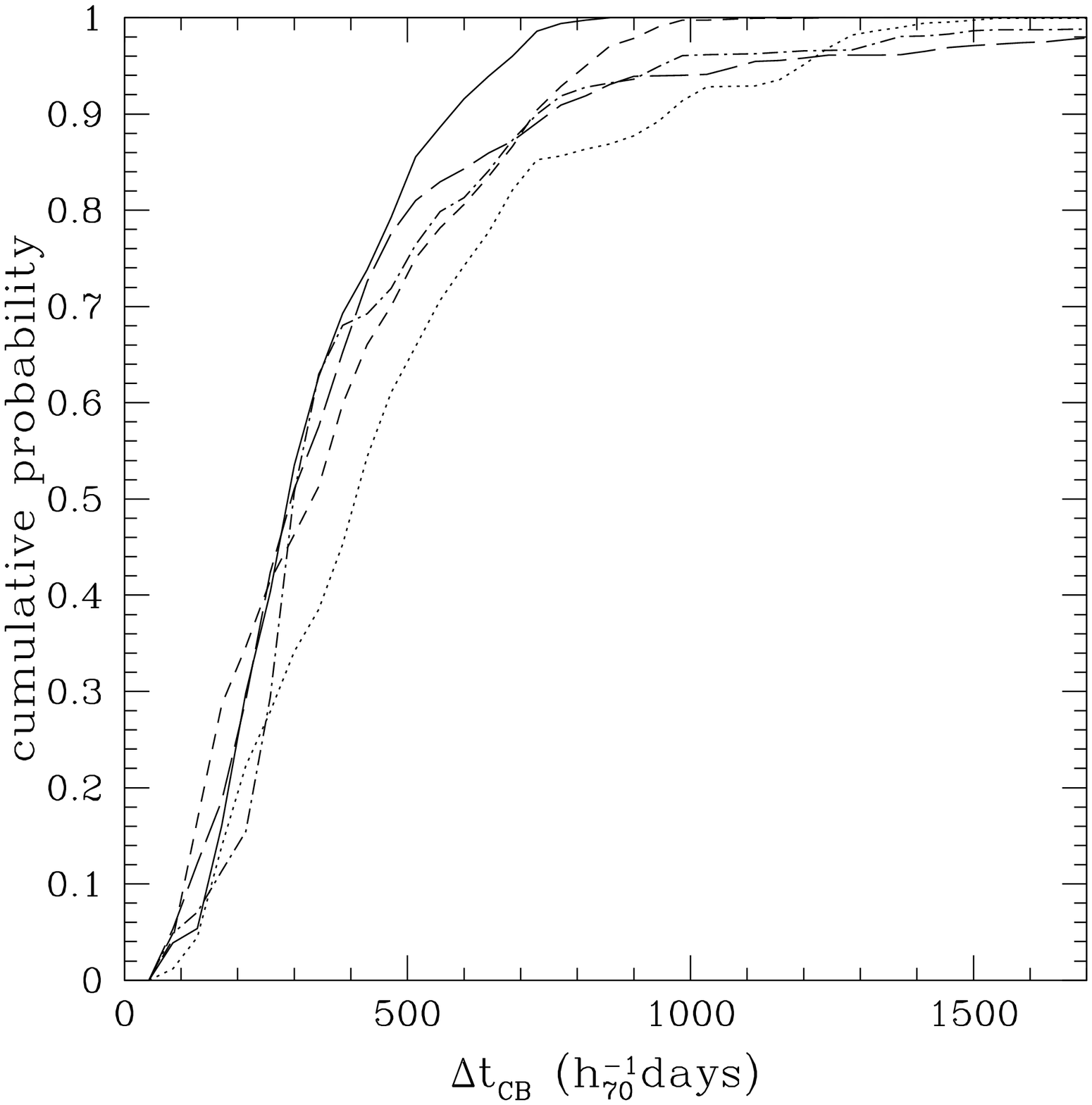}\\
    \FigureFile(65mm,65mm){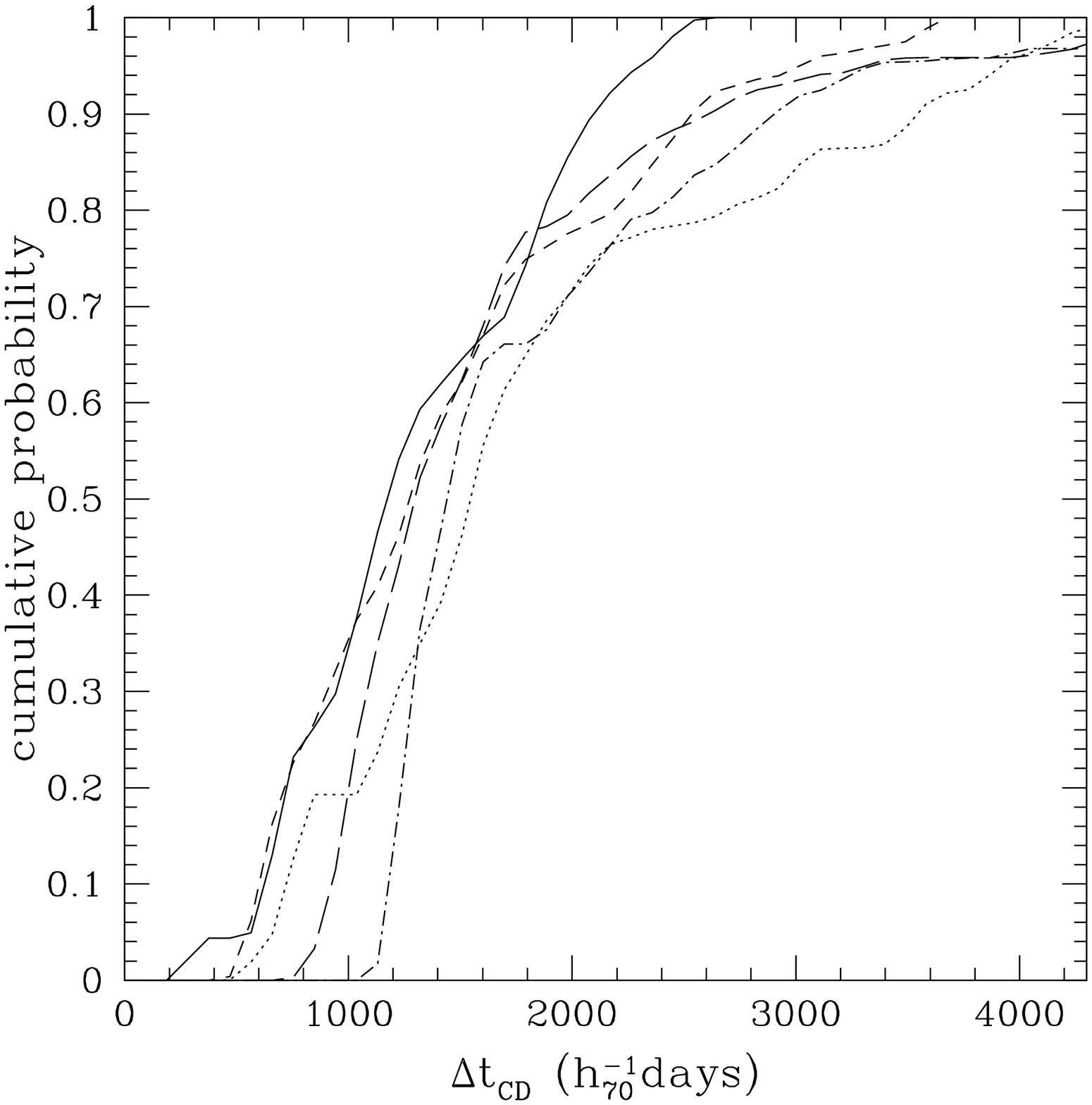}
    \FigureFile(65mm,65mm){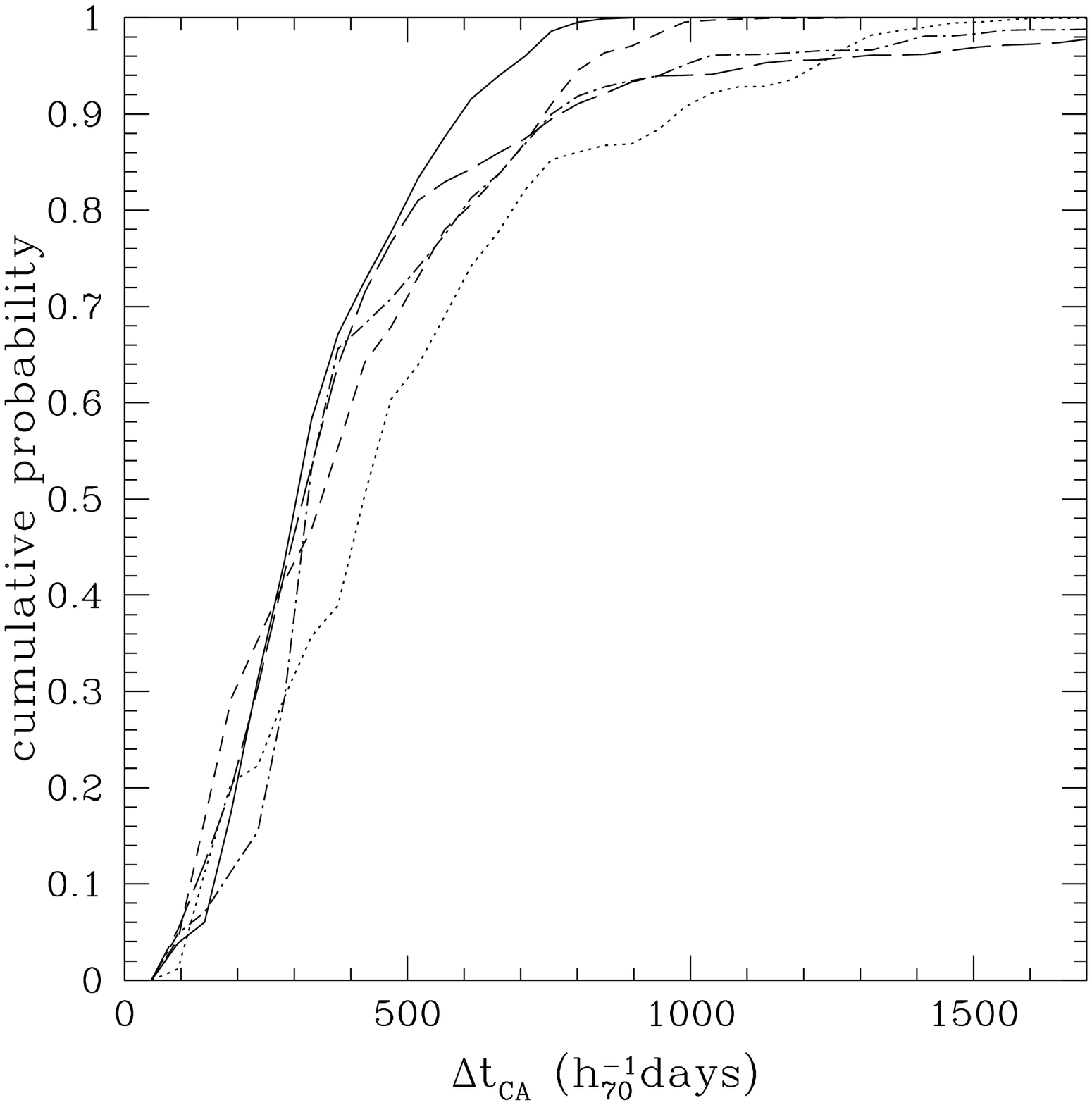}\\
    \FigureFile(65mm,65mm){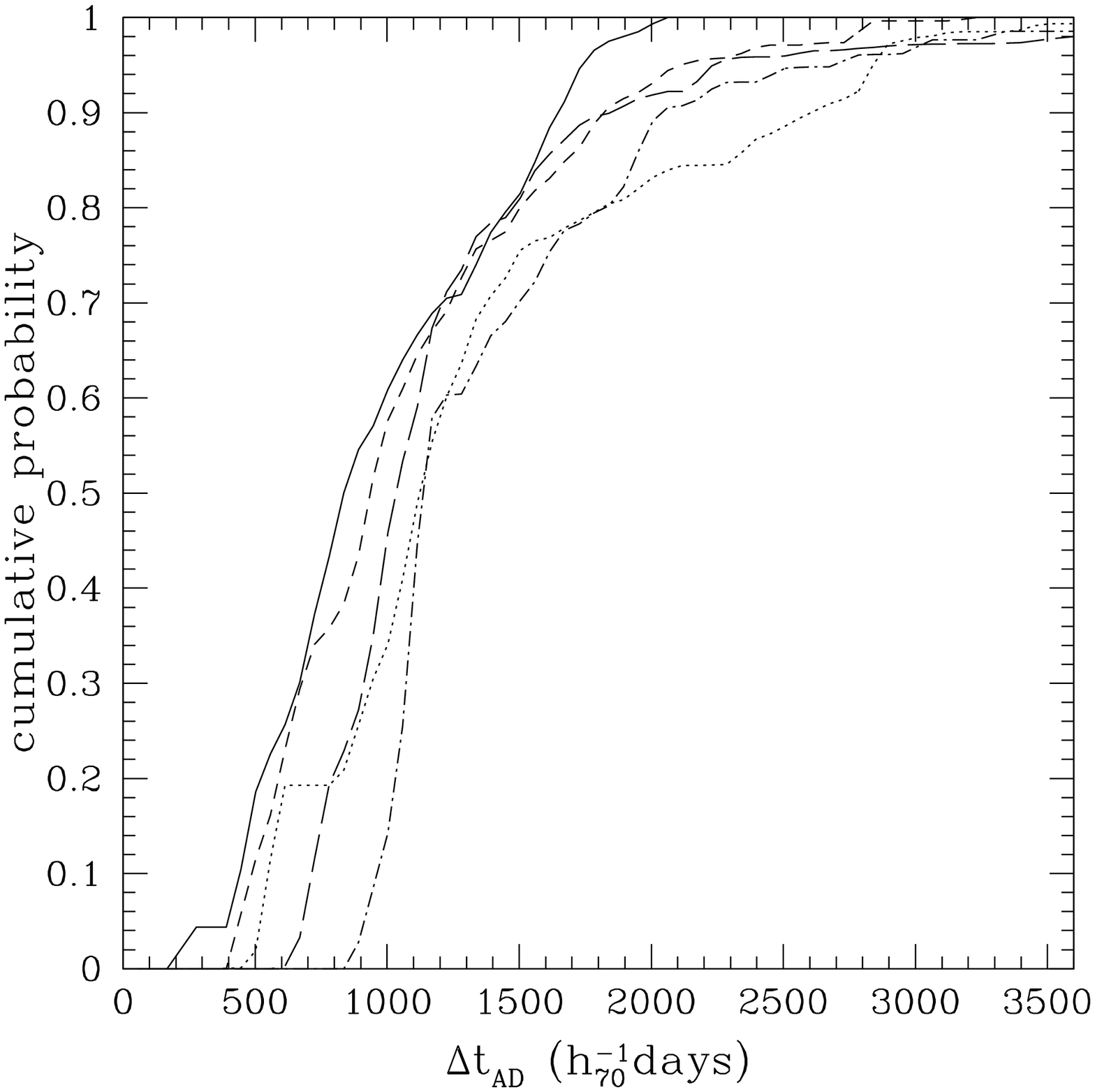}
    \FigureFile(65mm,65mm){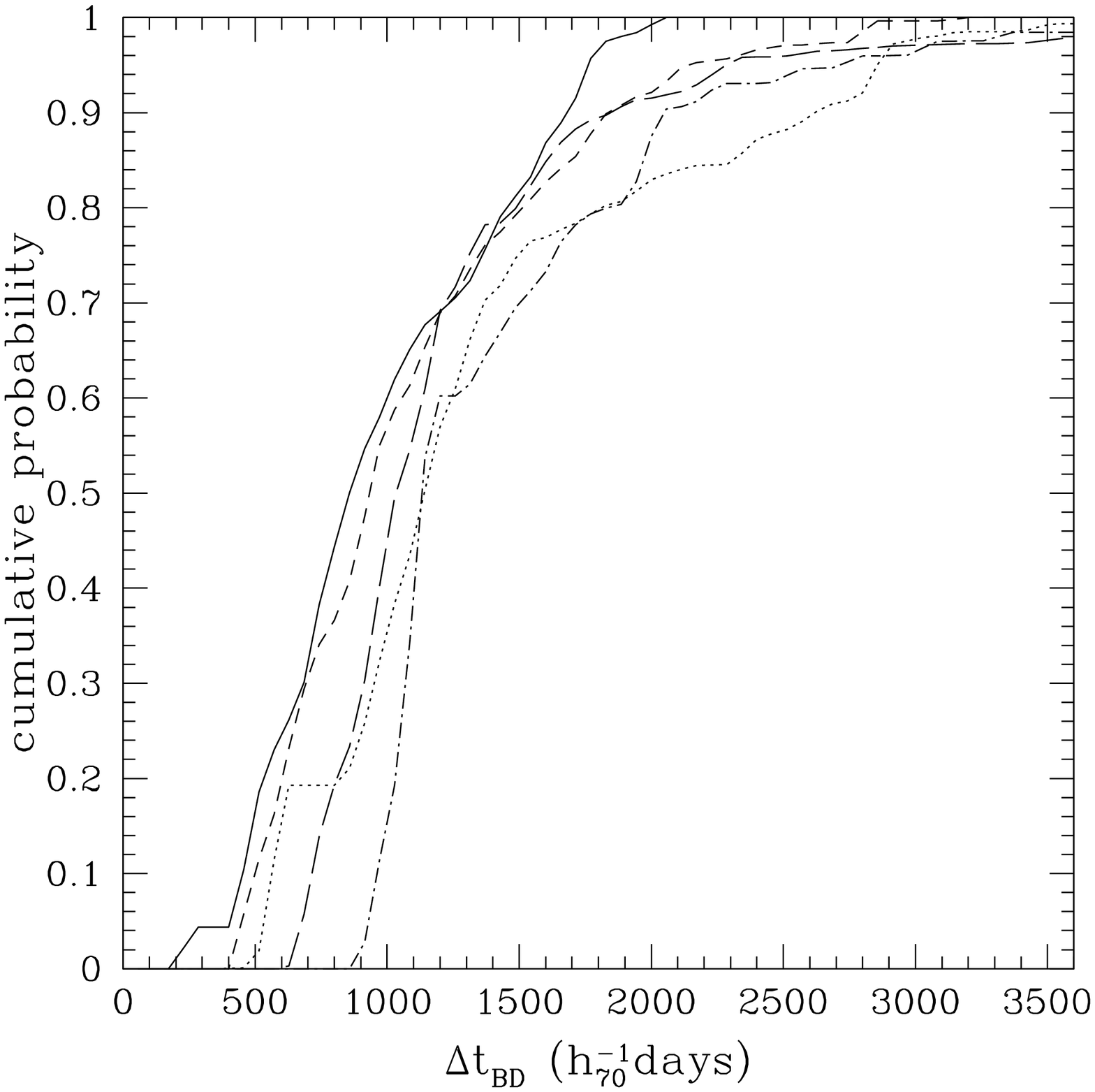}
  \end{center}
  \caption{Cumulative probability distribution functions of time
    delays of any pairs. In each panel, we show the PDFs for different
    inner slopes: $\alpha=0.5$ ({\it solid}), $0.75$ ({\it
    short dashed}),  $1.0$ ({\it dotted}), $1.25$ ({\it
    long dashed}), and $1.5$ ({\it  dot-dashed}).}
\label{cumulatedpdft_rs40}
\end{figure}

To evaluate distributions of predicted time delays, we construct the
probability distribution functions (PDFs) of all the six time delays 
($\Delta t_{\rm BA}$, $\Delta t_{\rm CB}$, $\Delta t_{\rm CD}$,
$\Delta t_{\rm CA}$, $\Delta t_{\rm AD}$, $\Delta t_{\rm BD}$) and the 
ratios between them, by summing up all acceptable models with the
weight of ${\rm exp}[-(\chi^2-\chi_{\rm min}^2)/2]$. Again, we neglect
the distributions of $\chi^2$ around each local minimum, since the 
predicted time delays does not significantly change around each local
minimum and the uncertainties of time delays are similar for different
local minima. Because of the fact that width of the valley of the local minimum
is not significantly changing, we can exclude the statistical weight.
Our choice of the maximum $\chi^2$, 11.8, is not
problematic because it corresponds to more than $99\%$ ($\Delta
\chi^2=11.3$) confidence limit even for ${\rm DOF}=3$. They enable one 
to show that observing the time delays may be able to constrain the
mass profile, especially the slope $\alpha$. The PDFs are constructed as
\begin{equation}
 P_{ij}(\Delta t|\alpha) \propto \sum_k^{100}
\exp \left( -\frac{\chi_k^2-\chi_{\rm min}^2}{2}
\right) \delta (\Delta t-\Delta t_{ij,k}),
\end{equation}
where subscript min denotes the minimum $\chi^2$ (for each fixed
$\alpha$). In fact, we divide each time delay into 100 cells, which
all have regular intervals, and calculate the PDF on each cell.
The PDFs in two and three time delay spaces are defined in the
same way. Similarly, the cumulative probability distribution
function can also be defined as
\begin{equation}
P_{ij}(<\Delta t|\alpha)=\int_0^{\Delta t} P_{ij}(\Delta t'|\alpha)
d(\Delta t').
\end{equation}

\begin{figure}
  \begin{center}
    \FigureFile(120mm,80mm){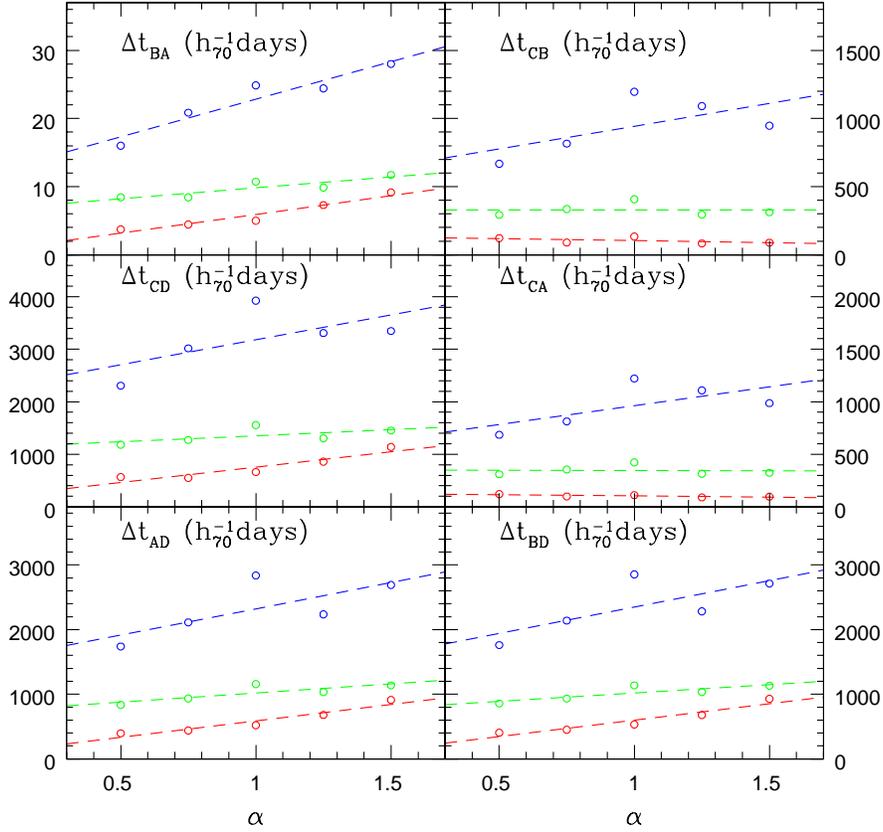}
  \end{center}
  \caption{Time delays at which the cumulative probabilities reach 5\%
    (red), 50\% (green), and 95\% (blue), are plotted as a function of
    the inner slope $\alpha$. The dashed line indicates the linear fit
    (eq. [\ref{eq:fit}]). The fitting parameters are summarized in
    Table \ref{tab:fit}.}
\label{alpha55095ptstij}
\end{figure}

Before going to the PDFs of time delays themselves, we see the PDFs of
time delay ratios in Figure \ref{ratiotpdf_rs40}. They slightly
broaden but clearly reveal the proportionality seen in
Figure \ref{tij_rs40al}. The PDFs of ratio
$\Delta t_{\rm CD}/\Delta t_{\rm BA}$ for $\alpha=1.0$ is in good
agreement with the value $143\pm16$ of \citet{oguri04}. However,
mean values of some ratios are changing significantly with changing
$\alpha$. For example,
the ratio $\Delta t_{\rm CD}/\Delta t_{\rm BA}$ for $\alpha=0.5$ and
$1.5$ are $\sim 155$ and $\sim 125$, respectively. The ratios $\Delta
t_{\rm AD}/\Delta t_{\rm BA}$ and $\Delta t_{\rm BD}/\Delta t_{\rm
  BA}$ depend weakly on the slope --- the value of $\alpha$ has
relatively much effect between image C and the other images. Thus,
from the measurements of the
time delay between A and B, we can estimate the time delays between A
and D, and B and D in good accuracy, regardless of the inner slope
$\alpha$. The dependence of image C on the slope $\alpha$ corresponds
the edges we have seen in Figure \ref{tij_rs40al}, to some extent. We
note that this figure is important to determine the observing
strategy of the time delays after measuring the shortest time delay
$\Delta t_{\rm BA}$. The figure also implies that determining multiple
time delays leads to tighter constraints on the inner mass slope
$\alpha$.  

Next we see the PDFs of time delays. Figure \ref{cumulatedpdft_rs40}
shows the cumulative probability of all the six time delays. The
probability has clear features which we cannot easily see in the
time delay space
(Figure \ref{tij_rs40al}), which are summarized as follows: 

1. The models with smaller $\alpha$ have a large fraction of short
time delays, and the shorter minimum and maximum time delays. For
instance, $\alpha=0.5$ model has predicted $\Delta t_{\rm BA}$ of
$\sim 1 - 26$ $h^{-1}_{70}$ days, and its cumulative probability reaches
$50\%$ at $\Delta t_{\rm BA}\sim 8$ $h^{-1}_{70}$ days which is nearly
equal to the minimum value of $\alpha=1.5$ model, while $\alpha=1.5$
model has predicted $\Delta t_{\rm BA}$ of $\sim 8 - 82$ $h^{-1}_{70}$
days, and its cumulative probability reaches $50\%$ at
$\Delta t_{\rm BA}\sim 11$ $h^{-1}_{70}$ days. 

2. The larger $\alpha$ model has a longer tail of the probability
distribution and hence the model uncertainties becomes larger with
increasing $\alpha$. 

3. For only $\Delta t_{\rm CB}$ and $\Delta t_{\rm CA}$ the probability
in the short time delay region gives different feature ---
$\alpha$ dependence cannot be almost seen in a region of the short time
delays. 

Put another way, increasing $\alpha$ makes the probability
distribution to shift to longer time delays and to have longer
tails. Thus, in principle the measurement of time delays will
constrain an acceptable range of $\alpha$. However, only one observed
time delay, for example $\Delta t_{\rm BA}$ which is the shortest one
and likely to be measured most easily, constrains $\alpha$ weakly,
given the large overlap of the PDFs seen in
Figure \ref{cumulatedpdft_rs40}. The exception is the observation of
very short or long time delay, which gives us the higher or lower
limit on the value of $\alpha$.  Because of the third feature, it is
supposed that constraint from the time delay between image B and C
is weak, if the time delay is relatively short.
 
\begin{table}
\begin{center}
\caption{Fitting Result}\label{tab:fit}
    \begin{tabular}{lcccccc}
\hline \hline
     $\Delta t_{ij}$ & $C_1$(95\%) & $C_2$(95\%) & $C_1$(50\%) &
$C_2$(50\%)& $C_1$(5\%) & $C_2$(5\%)\\
\hline
$\Delta t_{\rm BA}$ & 11.8 & 11.0 & 6.6 & 3.2 & 0.4 & 5.5\\
$\Delta t_{\rm CB}$ & 609.1 & 334.3 & 330.0 & $-1.7$ & 132.3 & $-29.1$\\
$\Delta t_{\rm CD}$ & 2232.3 & 949.1 & 1125.1 & 228.6 & 173.1 & 582.9\\
$\Delta t_{\rm CA}$ & 603.7 & 359.4 & 350.3 & $-4.6$ & 126.6 & $-22.9$\\
$\Delta t_{\rm AD}$ & 1512.0 & 810.9 & 739.1 & 280.6 & 79.1 & 510.3\\
$\Delta t_{\rm BD}$ & 1532.3 & 818.3 & 758.6 & 260.0 & 89.7 & 509.1\\
\hline
     \end{tabular}
\end{center}
\end{table}

As described above, the famous mass-slope versus $h_{70} \Delta t_{ij}$
degeneracy may explain that larger $\alpha$ models give longer time
delays. To test this more explicitly, Figure \ref{alpha55095ptstij}
shows time delays at which the cumulative PDFs reach 5\%, 50\%,
and 95\% for different values of the inner slope $\alpha$. We can
fit them with simple linear lines well.
Specifically, we fit them as
\begin{equation}
\Delta t_{ij}(P_{ij}(<\Delta t_{ij}|\alpha)=\mbox{5\%, 50\%, 95\%}) =
(C_1+C_2\alpha)h_{70}^{-1}{\rm days}.
\label{eq:fit}
\end{equation}
We summarize the fitting parameters $C_1$ and $C_2$ in Table
\ref{tab:fit}. But the time delays do not exactly obey the linear law
of the scaling relation $\propto (\alpha-1)$. While the discrepancy
may reflect the complicated mass distribution of the lensing cluster, 
we still have the qualitative relationship between the slope of the
mass distribution and time delays. 

\section{Constraints on the Radial Slope of the Cluster Mass
\label{sec:const}}
From the discussions in the previous section, it is not obvious how
well we can constrain the inner slope $\alpha$ from the measurements
of time delays. Thus, in this section we present a Monte-Carlo
analysis for this. First, we assume the value of the inner mass slope
$\alpha$ and randomly pick up a model from 100 models
which are used to construct the PDFs. In other word, we assume that
one, two, or three of time delays the model predicts are ``observed.''
The errors of the ``observed'' time delays are assumed to be $10\%$
Gaussian because most of observed time delays so far have errors of
$\sim10\%$. Then, we calculate the likelihood function  
\begin{equation}
{\cal L}(\alpha)=\int P_{ij}(\Delta t|\alpha) G(\Delta t_{ij})
d (\Delta t_{ij})
\label{eq:likelihood}
\end{equation}
for each $\alpha$. $G(\Delta t_{ij})$ is a Gaussian function
\begin{equation}
G(\Delta t_{ij})=\frac{1}{\sqrt{2\pi} \sigma_{\Delta t_{ij}}}
\exp \left[-\frac{(\Delta t_{ij}-\Delta t_{ij, 
{\rm obs}})^2}{2\sigma_{\Delta t_{ij}}^2}\right],
\label{eq:gauss}
\end{equation}
with $\Delta t_{ij, {\rm obs}}$ and $\sigma_{\Delta t_{ij}}=0.1\Delta
t_{ij, {\rm obs}}$ being the value of randomly generated ``observed''
time delay  and its error, respectively. Hereafter we consider the
three situations: 1) only one time delay ($\Delta t_{\rm BA}$) is
measured, 2) two of three time delays ($\Delta t_{\rm BA}$, $\Delta
t_{\rm CB}$) are measured, and 3) all the three time delays ($\Delta
t_{\rm BA}$, $\Delta t_{\rm CB}$, $\Delta t_{\rm CD}$) are measured.
For the cases of multiple measurements of time delays, the likelihood
becomes multiple-order integral. For each $\alpha$, we use nine models
which are randomly chosen from 100 fitted models. We calculate the
integral on each cell in same way as calculating PDFs. 
Table \ref{tab:input}
shows time delays in units of $h_{70}^{-1}$ days, which we adopt to
constrain the slope $\alpha$. For instance, when we assume that three
time delays of a model of (1.25-6), we insert $\Delta t_{{\rm BA},
{\rm obs}}=27.4$, $\Delta t_{{\rm CB}, {\rm obs}}=1353.4$, $\Delta
t_{{\rm CD},{\rm obs}}=4567.2$, $\sigma_{\Delta t_{\rm BA}}=2.74$,
$\sigma_{\Delta t_{\rm CB}}=135.34$, $\sigma_{\Delta t_{\rm CD}}=
456.72$ in units of $h_{70}^{-1}$days into equation (\ref{eq:gauss}),
and then from equation (\ref{eq:likelihood}) we calculate the
likelihood function for each $\alpha$ with the PDF constructed in \S
\ref{sec:predtij}. 

\begin{longtable}{lccc}
  \caption{Input Time Delays in Likelihood Analysis}\label{tab:input}
\hline \hline
Name & $\Delta t_{\rm BA}$ & $\Delta t_{\rm CB}$ & $\Delta t_{\rm CD}$\\
\hline
\endhead
\hline
\endfoot
\hline
\multicolumn{3}{l}{\hbox to 0pt{\parbox{180mm}{\footnotesize
  \footnotemark[$*$] Name of data are expressed as
(input $\alpha$-data number).\\
 \par\noindent
  \footnotemark[$\dagger$] Time delays in units of $h_{70}^{-1}$. The
  observational errors \\are assumed to be 10\%.
}}}
\endlastfoot
0.5-1 & 9.0 & 281.0 & 929.1 \\
0.5-2 & 11.4 & 591.3 & 1737.4 \\
0.5-3 & 6.7 & 274.7 & 966.7 \\
0.5-4 & 11.1 & 62.4 & 1783.3 \\
0.5-5 & 2.0 & 40.0 & 242.3 \\
0.5-6 & 14.9 & 661.6 & 2048.2 \\
0.5-7 & 2.8 & 65.7 & 489.2 \\
0.5-8 & 10.4 & 152.9 & 1173.0 \\
0.5-9 & 14.4 & 248.2 & 1910.4 \\
0.75-1 & 5.0 & 231.8 & 1151.2 \\
0.75-2 & 21.1 & 456.3 & 2482.0 \\
0.75-3 & 14.2 & 482.7 & 1539.8 \\
0.75-4 & 13.7 & 659.0 & 2443.8 \\
0.75-5 & 8.5 & 384.7 & 1435.5 \\
0.75-6 & 8.7 & 379.1 & 1327.6 \\
0.75-7 & 7.7 & 350.4 & 1257.7 \\
0.75-8 & 15.8 & 602.2 & 2191.1 \\
0.75-9 & 12.6 & 685.6 & 2293.0 \\
1.0-1 & 13.7 & 589.2 & 2240.6 \\
1.0-2 & 17.6 & 388.6 & 1188.9 \\
1.0-3 & 22.5 & 1211.2 & 3995.2 \\
1.0-4 & 7.9 & 254.2 & 1113.2 \\
1.0-5 & 5.0 & 133.6 & 688.0 \\
1.0-6 & 21.4 & 1154.7 & 3687.3 \\
1.0-7 & 11.0 & 475.8 & 1679.8 \\
1.0-8 & 9.4 & 391.7 & 1407.1 \\
1.0-9 & 5.7 & 67.0 & 617.4 \\
1.25-1 & 25.7 & 1490.8 & 4116.4\\
1.25-2 & 27.7 & 1509.39 & 4522.2\\
1.25-3 & 7.3 & 119.1 & 868.6\\
1.25-4 & 38.1 & 1190.5 & 4671.4 \\
1.25-5 & 11.1 & 172.1 & 1815.6 \\
1.25-6 & 27.4 & 1353.4 & 4567.2 \\
1.25-7 & 10.8 & 353.2 & 1629.7 \\
1.25-8 & 7.9 & 167.1 & 849.8 \\
1.25-9 & 9.9 & 254.3 & 1166.2 \\
1.5-1 & 18.9 & 872.0 & 2796.9 \\
1.5-2 & 22.8 & 1157.7 & 3635.5 \\
1.5-3 & 35.8 & 1877.9 & 5021.0 \\
1.5-4 & 11.1 & 286.7 & 2000.2 \\
1.5-5 & 10.8 & 254.6 & 1320.7 \\
1.5-6 & 15.9 & 653.5 & 2406.9 \\
1.5-7 & 9.4 & 133.1 & 1159.5 \\
1.5-8 & 8.3 & 17.2 & 1311.6 \\
1.5-9 & 40.0 & 1324.3 & 4331.2 \\
 & & &
\end{longtable}

\begin{figure}
  \begin{center}
    \FigureFile(120mm,80mm){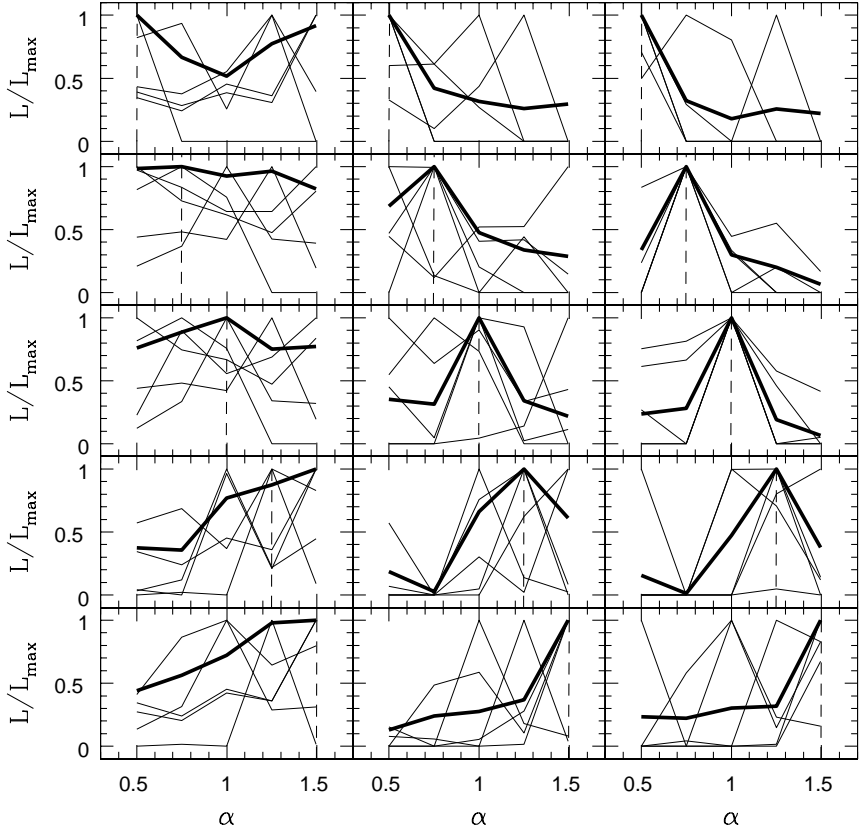}
  \end{center}
  \caption{The likelihood function ${\cal L}/{\cal L}_{\rm max}$ as a
    function of the inner slope $\alpha$. From top to bottom, the
    input $\alpha$, denoted by short dashed lines, are 0.5, 0.75, 1.0,
    1.25, 1.5, respectively. The numbers of time delays for
    constraining the slope are one ($\Delta t_{\rm BA}$), two ($\Delta
    t_{\rm BA}$, $\Delta t_{\rm CB}$), and three ($\Delta t_{\rm BA}$,
    $\Delta t_{\rm CB}$, $\Delta t_{\rm CD}$) from left to right
    panels. 
    Thin solid lines show the distributions of
    ${\cal L}/{\cal L}_{\rm max}$
    for each realizations (only 5 out of 9 realizations are shown,
    mainly because of the illustrative reason), and thick solid lines
    indicate the distributions after averaging over all 9 realizations.}
\label{allike}
\end{figure}

The results are shown in Figure \ref{allike}. Each thin and thick
solid line show likelihoods from each realization and likelihoods
averaged over nine realizations. From only one time delay
measurement, one cannot constrain the inner mass profile $\alpha$
significantly, as expected. But by measuring multiple time delays the
inner slope can be constrained with good accuracy
($\sigma_{\alpha}\lesssim0.25$). While each time delay, especially
$\Delta t_{\rm CB}$, gives us little information, 
the combination of time delays leads us to a true value of the slope.
We then conclude that longer time delays in addition to the shortest
time delay are necessary to obtain tight constraints on the inner
slope $\alpha$. 

\section{Discussion and Conclusions\label{sec:discon}}
We have presented predictions of time delays for the giant quadruple
lensed quasar SDSS J1004+4112, and investigated the relationship
between the time delays and the inner slope of the lensing
cluster. We adopt a two-component model in which the brightest cluster
galaxy and the cluster are described by the singular isothermal
ellipsoid and the generalized NFW profile. We parameterize the inner
slope of the cluster component by $\alpha$ such that $\rho\propto
r^{-\alpha}$ in the innermost region. The values of the slope $\alpha$
we have considered are 0.5, 0.75, 1.0, 1.25, and 1.5, while fixing the
scale length to be $40\arcsec$. We have derived a group of mass models
that fit the observables, and calculated the range of predicted time
delays for each $\alpha$. For observables, we have used the data of
image positions, fluxes, and the position angle of the galaxy to
predict the time delays. We have obtained a set of 100 acceptable
models for each $\alpha$, and constructed the PDFs of the predicted
time delays and the ratios between them.

We have found that predicted time delays indeed depend on the
inner slope, such that the steeper inner profiles (larger $\alpha$)
predict longer time delays. All the two of them are approximately
proportional to each other, but the ratios depend slightly on the
inner slope, which suggests that different $\alpha$ leads to a
different structure of the PDFs of multiple time delays. 
The larger $\alpha$ models predict longer minimum and maximum time
delays, and have longer tails at the higher end of the cumulative
probabilities. The time delays $\Delta t_{\rm CB}$ and $\Delta t_{\rm
 CA}$ give slightly different feature in short time delay --- the
models with different $\alpha$ show almost same distributions. It is
interesting that the model uncertainties resemble the inner slope
uncertainty in prediction of time delays, and it would be a clue to
investigate the model uncertainties.

To illustrate how well we can constrain the inner slope by adding
measurements of time delays, we have calculated the likelihood
function for $\alpha$. Figure \ref{allike} has shown that the
slope $\alpha$  is constrained weakly with the measurement of one time
delay. However, we also found that multiple time delays will results
in reasonably strong constraints ($\sigma_{\alpha}\lesssim0.25$) on the
inner slope. We note that we assumed observational errors of time
delays to be 10\%, which may be too conservative. For instance, Q
0957+561 has a time delay of $\sim 420$ days and the error is measured
to be 1\% \citep{kundic97,oscoz01}. If this level of errors can be
achieved for SDSS J1004+4112, we will be able to determine the inner
slope more tightly. 

Our model predictions offer useful guidance for photometric monitoring 
of this lens system to determine the time delays. 
We found that $95\%$ of models of $\alpha=0.5 - 1.5$ have the
predicted time delays of $\Delta t_{\rm BA}\lesssim 28$, $\Delta
t_{\rm CB}\lesssim 1400$, and $\Delta t_{\rm CD}\lesssim 3700$ in
units of $h_{70}^{-1}$ days. The first time delays is similar to those
in any other lensed quasars, and therefore it can be measured easily.
Indeed, the preliminary detection of the time delay between B and A
has been made to be $\sim 25$ days (C. S. Kochanek, private
communication). The second shortest time delay $\Delta
t_{\rm CB}$ is $\sim 30-40$ times larger than  $\Delta t_{\rm BA}$,
thus assuming the measured $\Delta t_{\rm BA}$ we predict $\Delta
t_{\rm CB}=750-1000$ days. This is somewhat longer than observed time
delays in any other systems, but is not impossible to measure. The
longest time delays need monitoring more than $\sim 10$ years, making
the measurement quite challenging. 

The mass profile of the lensing cluster can be constrained much better
if we combine the time delay measurements with other observations. 
For instance we may add constraints from multiply imaged galaxies
behind the cluster \citep{sharon05}. The multiple arcs that have
different redshifts (and therefore different Einstein ring radii) are
an independent tool to constrain the cluster mass profile, especially
the radial mass slope $\alpha$. Other such observations includes X-ray
measurements of intra-cluster medium and the measurement of the
velocity dispersion of the brightest cluster galaxy
(eq. [\ref{eq:bv}]). By combining these complementary information, we
will be able to reveal the detailed distribution of the dark matter of
the lensing cluster in a robust manner. 

\bigskip
We are grateful to Takahiko Matsubara, Naohisa Inada, and Chris
Kochanek for discussions.

\end{document}